\documentclass[journal=jacsat,manuscript=article]{achemso}

\usepackage{chemformula} 
\usepackage[T1]{fontenc} 

\author{Xuance Jiang}
\email{xuance@ucsb.edu}
\affiliation{Department of Chemistry and Biochemistry, University of California, Santa Barbara, Santa Barbara, California 93106, United States}

\author{Vojtech Vlcek}
\affiliation{Department of Chemistry and Biochemistry, University of California, Santa Barbara, Santa Barbara, California 93106, United States\\ Materials Department, University of California, Santa Barbara, Santa Barbara, California 93106}

\title[GW]
  {Cost Reduction in Spin-dependent Stochastic $GW$ Calculations}

\abbreviations{IR,NMR,UV}
\keywords{American Chemical Society, \LaTeX}

\begin{document}


\begin{abstract}
 We extend the stochastic $GW$ (s$GW$) formalism to fully spin-polarized systems, encompassing both collinear and non-collinear spin configurations. For non-collinear systems—where Kohn–Sham states are complex two-component spinors—we develop a complex-valued stochastic basis that preserves the real-valued external stochastic charge applied at time zero. This basis enables an unbiased evaluation of the random-phase approximation (RPA) screened interaction for spinors. Through error analysis and tests on real materials, we show that the performance of collinear s$GW$ retains the same time complexity as the spin-unpolarized s$GW$. The non-collinear s$GW$ incurs a computational cost two to three times higher than the spin-unpolarized version, while preserving linear scaling with low multiplicity. By unifying collinear and non-collinear treatments within a single scalable framework, our work paves the way for routine many-body predictions in large scale magnetic and spin–orbit–coupled material systems.
\end{abstract}


\maketitle

\section{Introduction}
Many-body perturbation theory $GW$ approximation remains the gold-standard for predicting quasiparticle (QP) energies in weakly correlated solids and molecules~\cite{hedin1965new,strinati1980dynamical,strinati1982dynamical,Hybertsen1986,Umari2009,Umari2010,Deslippe2012,Ljungberg2015,Govoni2015,Wilhelm2018}. By dressing the one-electron Green’s function $G$ with a dynamically screened Coulomb interaction $W$~\cite{hedin1965new,martin2016interacting,aryasetiawan1998gw}, the method captures exchange and long-range correlation beyond density-functional theory, routinely achieving high accuracy for band gaps, ionization energies, and electron affinities~\cite{Hybertsen1986,shishkin2007self,van2015gw,golze2019gw}. However, the scaling of $GW$ method is steep~\cite{Umari2010,Govoni2015}. 

Stochastic $GW$ (s$GW$) was introduced to overcome the steep cost of conventional $GW$ implementations by replacing explicit summations over empty states with random sampling~\cite{Neuhauser2014,Vlcek2017,vlvcek2018swift,Vlcek2018PRM}. In this framework, Green’s functions and $W$ are evaluated with stochastic orbitals, reducing the scaling to nearly linear in system size while retaining controllable statistical error bars.

Although some deterministic $GW$ calculations have been implemented~\cite{sakuma2011gw,scherpelz2016implementation,barker2022spinor}, most s$GW$ studies have focused on non-magnetic or collinearly spin-polarised systems, leaving large-scale materials with non-collinear spins or strong spin–orbit coupling beyond reach. Such systems—ranging from skyrmionic magnets to topological insulators—require a spinor treatment in which wavefunctions are complex two-component objects.

Here we develop and test the spin-resolved s$GW$ that unifies collinear and fully non-collinear spin configurations within a single stochastic framework. For collinear magnets, we retain separate stochastic sampling of majority and minority channels. For non-collinear cases, we introduce a complex stochastic orbital basis that preserves the real-value stochastic external charge applied at time zero under time-dependent Hartree (TDH) propagation~\cite{hedin1999correlation,martin2016interacting}. This basis enables an unbiased evaluation of the random-phase-approximation effective screened interaction
for spinors. By incorporating collinear and fully non-collinear spin configurations into a single, scalable scheme, our approach makes many-body quasiparticle (QP) calculations practical for large magnetic and spin–orbit–rich materials.

The paper is organized as follows. We overview the stochastic $GW$ on spinor in Sec. II. In Sec. III we compare the performance of no spin and spin resolved s$GW$ by analyzing the time cost of the most time-consuming stochastic TDH. Then we discuss the test on collinear and non-collinear systems, followed by a conclusion.

\section{Methodology}

As the implementation is based on earlier works~\cite{Neuhauser2014,Vlcek2017,vlvcek2018swift,Vlcek2018PRM}, we will here review only the key novel aspects of the computational methodology. The key is the recasting of the operator actions as sampling via random operators. This is accomplished in real space by introducing a random sampling of real-space coordinates and spin channels, with a random sign at each grid point 
\begin{equation}
    \bar{\zeta}(\mathbf{r}\sigma) = \pm (dV)^{-\frac{1}{2}},
\end{equation}
where $dV$ is the grid volume element. They fulfill the orthonormal relation $\{\bar{\zeta}(\mathbf{r}\sigma) \bar{\zeta}(\mathbf{r}'\sigma')\} = (dV)^{-1} \delta_{\mathbf{r}, \mathbf{r}'}\delta_{\sigma\sigma'}$, where $\delta_{\mathbf{r}, \mathbf{r}'},\delta_{\sigma\sigma'}$ are Kronecker delta and we use \{...\} to refer to a statistical average. The stochastic resolution of identity (sRI) corresponds to $\mathcal{I} = \{ |\bar{\zeta}\rangle \langle \bar{\zeta}|\}
$ in the chosen space-spin basis. Note that this is not the only choice of random function set. For any arbitrary phase function $f(\mathbf{r}\sigma)$, $f(\mathbf{r}\sigma)\bar{\zeta}(\mathbf{r}\sigma)$ also works as a good random functions set. We will show that the proper choice of $f$ can help circumvent the complex source charge problem.

\subsection{Green's function sampling}

The sampling of the non-interacting Kohn-Sham Green’s function is the key component of the stochastic perturbative treatment (e.g., s$GW$)~\cite{Neuhauser2014,Vlcek2017,vlvcek2018swift,Vlcek2018PRM}. Instead of representing $G$ via its individual orbital (basis) components, we express it as a statistical average over random functions, here generalized to spinors. Formally, 
\begin{equation}
i\hat{G}_0(t) = e^{-i \hat{H}_0 t} \left[(\hat{I} - \hat{P})\theta(t) - \hat{P} \theta(-t)\right],    
\end{equation}
where $\hat{P}$ is the projection operator onto the space of occupied states $\hat{P} = \sum_{n \leq N_{\text{occ}},\sigma=1,2} |\phi_{n\sigma}\rangle \langle \phi_{n\sigma}|$. In the spin-dependent $GW$ framework, the screened interaction $W$ is spin-diagonal and therefore does not generate spin-flip terms. However, the SOC leads to the off-diagonal term in $\hat H_0$ and thus the time evolution operator $e^{-i \hat{H}_0 t}$ will mix the channels due to the block-off-diagonal terms. Therefore, the spin off-diagonal self-energy contributions arise solely from the Green’s function $G$ and the self-energy naturally acquires spin off-diagonal components even though $W$ itself acts identically on both spin channels.

Introducing sRI using the random basis into $G$, we get:
\begin{equation}
iG(\mathbf{r}\sigma, \mathbf{r}'\sigma', t) = \{\zeta(\mathbf{r}\sigma, t)\, \bar{\zeta}(\mathbf{r}'\sigma')\},    
\end{equation}
where $|\zeta(r\sigma,t)\rangle \equiv i\hat {G}(t) |\bar{\zeta}(r\sigma)\rangle$, in which we employ the split-operator (Trotter) time propagation method as in~\cite{trotter1959product}.
This equation is central to the stochastic $GW$ method, as it rewrites the Green’s function as a sum of separable terms involving random functions.

Within this formalism, the time-dependent expectation value of the self-energy is 
\begin{equation}
\langle \phi_{\sigma} | \hat\Sigma(t) | \phi_{\sigma'} \rangle = \Biggl\{ \int \phi_{\sigma}(\mathbf{r})\, \zeta(\mathbf{r}\sigma, t)\, u(\mathbf{r}\sigma', t) \, d\mathbf{r} \Biggr\},   
\end{equation}
with a time-ordered potential representing the density fluctuation induced by the addition/removal of a particle to/from state $\phi_{\sigma'} (r')$ 
\begin{equation}\label{eq:to_pot_W}
u(\mathbf{r}\sigma', t) = \int W(\mathbf{r}, \mathbf{r}', t)\, \bar{\zeta}(\mathbf{r}'\sigma')\, \phi_{\sigma'}(\mathbf{r}') \, d\mathbf{r}'.    
\end{equation}
Here, $W$ is the screened Coulomb interaction. In the random-phase approximation (consistent with the $GW$ approximation~\cite{vlcek2019stochastic,mejuto2022self}, the induced potential is described by the fluctuations of the total charge density (computed from the time-dependent Hartree calculations detailed below)\cite{hedin1999correlation,martin2016interacting}.  As such, $u(r\sigma', t)$ affects both spin channels identically and depends only on the perturbing state  $\phi_{\sigma'}(r')$. 



\subsection{Real-time evaluation of screened interactions}

The screened Coulomb interaction is conventionally split between the (dominant) exchange term computed explicitly and the polarization part due to the induced charge fluctuations. We will detail the calculations of the latter. Here, the time-ordered potential, which enters Eq.~\ref{eq:to_pot_W}, is obtained by a simple transformation (detailed in~\cite{vlvcek2018swift}) of its retarded counterpart polarization potential:
\begin{equation}
   u^R(\mathbf{r}, t) = \sum_{\sigma'}\int W^R(\mathbf{r}, \mathbf{r}', t)\, \bar{\zeta}(\mathbf{r}'\sigma')\, \phi_{\sigma'}(\mathbf{r}') \, d\mathbf{r}'. 
\end{equation}
This term is computed from time evolution of the charge density as $W$ is obtained as the potential due to the charge density fluctuations induced by the addition/removal of the electron. The details are provided in ~\cite{vlvcek2018swift}.


If the wavefunction $\phi_{\sigma'}$ can be expressed as a real-valued function of the position, we can follow the procedure applied in the non-spin-polarized s$GW$. Specifically, the external perturbation is now the product:
\begin{equation}
v_{\text{pert}}(\mathbf{r}) = \sum_{\sigma'}\int \nu(\mathbf{r}, \mathbf{r}')\, \bar{\zeta}(\mathbf{r}'\sigma')\, \phi_{\sigma'}(r')\, d\mathbf{r}',
\end{equation}
where $\nu(\mathbf{r}, \mathbf{r}') = |\mathbf{r} - \mathbf{r}'|^{-1}$ is the bare Coulomb interaction.

We perturb all occupied Kohn-Sham orbitals at time $t = 0$:
\begin{equation}
\phi_{n\sigma}^\lambda(\mathbf{r}, 0) = e^{-i \lambda v_{\text{pert}}(\mathbf{r})} \phi_{n\sigma}(\mathbf{r}), \quad n \leq N_{\text{occ}},
\end{equation}
using a small parameter $\lambda$ (typically $10^{-4}\,E_h^{-1}$).
The perturbed orbitals are propagated in time under the TDH Hamiltonian:
\begin{equation}
H^\lambda(t) = H_0 + v_H^\lambda(\mathbf{r}, t) - v_H(\mathbf{r}),
\end{equation}
where $v_H(\mathbf{r})=v_{H}[n](\mathbf{r}) = \int \nu(\mathbf{r}, \mathbf{r}') n(\mathbf{r}') \, d\mathbf{r}' \quad$ and $v_H^\lambda(\mathbf{r}, t)=v_{H}[n^\lambda](\mathbf{r})$ is the Hartree potential constructed from the time-dependent density:
\begin{equation}
n^\lambda(\mathbf{r}, t) =  \sum_{n \leq N_{\text{occ}},\sigma=1,2} |\phi_{n\sigma}^\lambda(\mathbf{r}, t)|^2.
\end{equation}

Except for the small systems, the total density fluctuation is sampled (rather than constructed by a sum over states). Hence, we first generate a set of $N_\eta$ stochastic occupied orbitals of the form:
\begin{equation}\label{eq:stoch_orb}
\eta^{\sigma}_l(\mathbf{r}) = \sum_{n \leq N_{\text{occ}}} \eta_{nl} \phi_{n\sigma}(\mathbf{r}), \quad l = 1, \dots, N_\eta, \sigma = 1, 2
\end{equation}
where the coefficients $\eta_{nl}$ are chosen randomly (e.g., with a random complex phase and unit magnitude).

Each stochastic orbital is perturbed as:
\begin{equation}
\eta_l^\lambda(\mathbf{r}, 0) = e^{-i \lambda v_{\text{pert}}(\mathbf{r})} \eta_l(\mathbf{r}),
\end{equation}
and propagated using the same time-dependent Hamiltonian $H^\lambda(t)$ as in the deterministic case.

The time-dependent density is computed as:
\begin{equation}
n^\lambda(\mathbf{r}, t) = C_{\text{norm}} \cdot \frac{1}{N_\eta} \sum_{l=1}^{N_\eta} \sum_\sigma|\eta_{l\sigma}^\lambda(\mathbf{r}, t)|^2,
\end{equation}
where $C_{\text{norm}}$ ensures the correct normalization (i.e., $\int n^\lambda(\mathbf{r}, t)\, d\mathbf{r} = N_e$).

The retarded potential $u_R(r, t)$ is then given by the linear response:
\begin{equation}
u_R(\mathbf{r}, t) = \frac{v_H^\lambda(\mathbf{r}, t) - v_H(\mathbf{r})}{\lambda}.
\end{equation}

The situation is different for spin noncollinear case (e.g., in the presence of SOC) as $\phi_{\sigma'}$ is generally a complex-valued function. As a result, the source charge and $u^R$ are complex, and we implement a $\phi$ dependent stochastic basis to solve for the induced potential:
To make the source charge $\bar{\zeta}\phi$ real and hence speed up the calculations, we let the new basis $\bar{\zeta}'$ be $\phi$ dependent. The new basis is,

\begin{equation}
    \bar{\zeta}'(\mathbf{r}\sigma)=\frac{\bar{\zeta}(\mathbf{r}\sigma)\phi_\sigma^*(\mathbf{r})}{|\phi_\sigma(\mathbf{r})|}
\end{equation}.

We can also check that this new basis satisfies $\{\bar{\zeta}'^*(\mathbf{r}\sigma) \bar{\zeta}'(\mathbf{r}'\sigma')\} = (dV)^{-1} \delta_{\mathbf{r}}$. Then the new source charge is $\bar{\zeta}'(\mathbf{r})\phi(\mathbf{r})=\bar{\zeta}(\mathbf{r})|\phi(\mathbf{r})|$.  This greatly simplifies the approach and it is possible to employ the same (TDH) approach as for the spin-independent evolution used in the past. Note that the additional complex phase $\phi/|\phi|$ is trivially absorbed in the green function $G$. An alternative is to compute the real and imaginary parts of the charge response separately~\cite{supp}, but this is significantly more computationally demanding. We therefore implement and test the $\phi$-dependent stochastic-basis method for the remainder of the paper.

\section{Convergence of the real-time stochastic formalism}


Deterministic $GW$ calculations require explicit construction and manipulation of large Green’s function and screened interaction matrices, with computational cost that grows steeply ($\mathcal{O}(N^{4})$) with system size~\cite{Umari2010,Govoni2015}. By contrast, the s$GW$ formalism avoids this bottleneck: the cost scales nearly linearly with the number of electrons~\cite{Neuhauser2014,Vlcek2017,vlvcek2018swift}, since operator actions are estimated through random sampling rather than full matrix operations. As we show below, the advantages of random sampling prevail even when considering systems with broken spin degeneracies, and, in fact, the stochastic formalism has a further significant reduction in the computational cost compared to the conventional (deterministic) calculations.

To quantify the scaling of the spin-resolved stochastic scheme, we focus on the dominant computational step: the real-time propagation of the electronic states, which are practically sampled by random vectors. To analyze the code performance, we estimate the time-cost ratios for non-spin, spin-collinear, and spin-noncollinear s$GW$. We find that for the most time-consuming step, the linear scaling in the non-spin polarized s$GW$ are preserve. Moreover, because of self-averaging in s$GW$, spin-resolved calculations require fewer random functions $N_\eta$ for sampling, leading to only a modest increase in time cost compared to non-spin-polarized s$GW$ (Table~\ref{tab:perform}).  In contrast, the conventional $GW$ implementations (in the energy domain), the self-energy calculation normally takes four times longer due to double the number of bands and double the size of each wave function~\cite{barker2022spinor,forster2023two}.  This is governed by the size of the operator matrices and the corresponding vectors. The spinor Hamiltonian $H^{\sigma_1\sigma_2}_{ij}$ and self energy $\Sigma(\omega)^{\sigma_1\sigma_2}_{ij}$ matrices are 4 times larger with spinor $\left | \phi^\sigma_j\right \rangle$ basis. Note that this is irrespective of the underlying physical interactions, i.e., if the calculation is performed in the enlarged basis, this leads to redundant computations which add to the total cost of the calculation.


First, we discuss the formalism employing random vectors spanning the space of single quasiparticle states. This stochastic formulation samples operator actions and turns expectation values into statistical estimators. The computational cost is thus governed by the variance of the sampling that determines the number of random vectors (samples) necessary to converge the expectation values to a predetermined threshold. The convergence threshold can be chosen according to the desired resolution of spectral features or the level of accuracy needed for quasiparticle energies.



We consider the time-dependent density obtained from propagated perturbed orbitals. In the first step, 
\begin{equation}
    \phi_{n\sigma}^{\lambda}(\mathbf{r},0+dt)=e^{-iH(0) dt}e^{-i\lambda v_{pert}}\phi_{n\sigma}(\mathbf{r}).
\end{equation}
As the random expansion coefficients $\eta$ mixing the eigenstates in Eq.~\ref{eq:stoch_orb} are time invariant, the random sample of occupied states is: 
\begin{equation}
    \eta_l^{\lambda}(\mathbf{r},0+dt)=\sum_{n \leq N_{\text{occ}}} \eta_{nl} \phi_{n\sigma}(\mathbf{r},0+dt),
\end{equation}
Note that the stochastic error in the next time step will accumulate on the new Hamiltonian. So the final stochastic error is not necessarily bounded and may lead to bias. But in the converged case, we assume that the time evolution of $\phi$ is achieved with an unbiased time-dependent Hamiltonian within a finite pattern that is governed by total time steps in our code (this is guaranteed by taking a large enough number of sampling vectors to reconstruct the density that enters $H$). So we can consider the total stochastic error as a fixed monotonic function of the stochastic error from the first step of propagation.  In the results section (Section IV), we indeed demonstrate that this is the case and that we achieve this expected behavior with a small number of random vectors. Then we consider the first time step below. The time-dependent density is,

\begin{equation}
\begin{split}
n^{\lambda}&(\mathbf{r},0+\mathrm{d}t)
= \frac{1}{N_\eta} \sum_{l=1}^{N_\eta} \sum_\sigma \left(
    \sum_{n\le N_{\mathrm{occ}}} |\eta_{nl}|^2 |\phi^{\lambda}_{n\sigma}(\mathbf{r},0+\mathrm{d}t)|^2 \right.\\
&\qquad \left. + 2\sum_{j<k}Re[\eta^{*}_{jl}\eta_{kl}\phi^{\lambda *}_{j\sigma}(\mathbf{r},0+\mathrm{d}t)\phi^{\lambda}_{k\sigma}(\mathbf{r},0+\mathrm{d}t)] \right).
\end{split}
\end{equation}

Since $|\eta_{nl}|=1$, the difference between the stochastic and deterministic density at the first time step is (we neglect $0+\mathrm{d}t$ below),

\begin{equation}
\begin{split}
\Delta n(\mathbf r)
&= \frac{2}{N_\eta} \sum_{l=1}^{N_\eta} \sum_\sigma \sum_{j<k}
   \cos\!\big(\theta_{jl}-\theta_{kl}
              +\theta^{\lambda }_{j\sigma}(\mathbf r)
              -\theta^{\lambda }_{k\sigma}(\mathbf r)\big) \\
&\quad \times |\phi^{\lambda }_{j\sigma}(\mathbf r)|\,|\phi^{\lambda }_{k\sigma}(\mathbf r)|.
\end{split}
\end{equation}

where we assume $\eta_{jl}=e^{i\theta_{jl}}$, $\phi^{\lambda }_{j\sigma}(\mathbf{r})=|\phi^{\lambda }_{j\sigma}(\mathbf{r})|e^{i\theta^{\lambda }_{j\sigma}(\mathbf{r})}$. Therefore, we can estimate the stochastic error of the first time step by calculating the variation of $\Delta n(\mathbf r)$ as,

\begin{equation}
\begin{split}
\mathrm{Var}(n^\lambda)=\mathrm{Var}(\Delta n(\mathbf r))
&= \frac{2}{N_\eta}  \sum_\sigma \sum_{j<k}
 (|\phi^{\lambda }_{j\sigma}(\mathbf r)|\,|\phi^{\lambda }_{k\sigma}(\mathbf r)|)^2.
\end{split}
\end{equation}

Then we analyze the error of the stochastic density in three cases with increasing complexity of the spin interactions.

\paragraph{Spin degenerate systems} For completeness, we first review systems that are fully spin symmetric. Since the computational time of the $GW$ calculations is dominated by the real-time TDH calculation, we analyze the number of stochastic orbitals $N_\eta$ required to converge the evolution of the density fluctuations. The effect of TRS reducing $N_\eta$ can be understood in a simple way when the SOC is negligible, and the two spin channels are always separated. TRS pairs averaging each other with half the $N_{\eta}$ required to converge the stochastic TDH. The $N_{\eta}/2$ spinor stochastical function $\eta^{\sigma}_l$ are effectively  $N_{\eta}$ non-spin polarized stochastical functions. Therefore, only half the spinor stochastic functions are needed to have a similar screening effect compared to the non-spin-polarized case. 

To confirm this point, we analyze the stochastic error of the density. When the system has TRS and no SOC, the spinors reduce to a spinless wavefunction and the variation of the first step density is,
\begin{equation}
\mathrm{Var}(n^{\lambda})=\frac{4}{N_\eta}\sum_{j<k}(|\phi^{\lambda }_{j}||\phi^{\lambda}_{k}|)^2.\end{equation}

Therefore, the relative density error that should be controlled can be written as the product of terms, 
\begin{equation}
\begin{split}
\frac{1}{n^\lambda}\sqrt{\mathrm{Var}(n^\lambda)}&=\sqrt{\frac{2\sum_{j<k}(|\phi^{\lambda }_{j}||\phi^{\lambda}_{k}|)^2}{N_\eta(\sum_{j}|\phi^{\lambda}_j|^2)^2}}\\
&\leq \sqrt{\frac{2(N_{occ}-1)} {N_{occ}N_\eta } }.
\end{split}
\end{equation}
Where $N_{occ}$ is the number of occupied states. As the system size increases, their values are kept bounded at $\sqrt{2/N_\eta}$. So only a small constant $N_\eta$ is required to achieve a small relative time-dependent density and thus a converged self-energy calculation. It has been illustrated in Refs.~\cite{vlvcek2018swift,Vlcek2018PRM}.

\paragraph{Collinear magnetic systems}

In this case, the spin channels are separate and preserve the spin $U(1)$ symmetry. The variation of density,

\begin{equation}
\mathrm{Var}(n^{\lambda})=\frac{2}{N_\eta}\left(\sum_{j<k}(|\phi^{\lambda }_{j\uparrow}||\phi^{\lambda}_{k\uparrow}|)^2+\sum_{j'<k'}(|\phi^{\lambda }_{j'\downarrow}\phi^{\lambda}_{k'\downarrow}|)^2\right).
\end{equation}

When the occupation numbers of spin-up and spin-down states are equal, as in antiferromagnetic systems, the variation is comparable to that of the non-spin-polarized case. Only when the two spin channels have unequal occupations does the variation become larger than in the non-spin case. In the extreme limit where a single spin channel is fully occupied with $2N_{occ}$ electrons while the other is completely empty, the variation scales as $2N_{occ}^2 - N_{occ}$ $,\mathrm{Re}[\phi^{\lambda *}{j'\uparrow}\phi^{\lambda}{k'\uparrow}]^2$, rather than $N_{occ}^2 - N_{occ}$ $,\mathrm{Re}[\phi^{\lambda *}{j'\sigma}\phi^{\lambda}{k'\sigma}]^2$ for the spin-degenerate case. In this situation, the number of stochastic orbitals $N_\eta$ must be doubled to bound the density variation at the same level as in the unpolarized case. Thus, only in the extreme case of complete spin polarization does the computational cost double relative to the spin-unpolarized case.

\paragraph{Systems with noncollinear magnetism}

The computational complexity is directly related to the preservation of the time reversal symmetry. If TRS is present, the perturbed states form Kramer pairs because the perturbation and propagation both preserve TRS. Therefore, the variation is,
\begin{align*}    
\mathrm{Var}(n^{\lambda})
&=\frac{2}{N_\eta}\left(\sum_{j<k}(\sum_\sigma |\phi^{\lambda }_{j\sigma}||\phi^{\lambda}_{k\sigma}|)^2\right)\\
&\sim \frac{4}{N_\eta}\left(\sum_\sigma \sum_{j<k}(|\phi^{\lambda }_{j\sigma}||\phi^{\lambda}_{k\sigma}|)^2\right), 
\end{align*}
where $j,k$ indices are from two TRS subspaces. So the variation is still comparable to the no-spin case with the same $N_\eta$. However, when the TRS is broken, the variation can increase up to twice the TRS preserved case, which indicates more $N_\eta$ ($1\sim 2$ $\times$ no spin case) are required to converge the density and the self-energy calculation. 

Note that the stochastic time-dependent Hartree (TDH) propagation cost scales linearly with both $N_\eta$ and the size of $\phi$. This scaling arises because each stochastic orbital must be propagated independently, and the computational effort for each propagation grows with the orbital size. For non-spin and collinear spin cases, the $\phi$ size is half that of the noncollinear case (Table~\ref{tab:perform}). Combined with the earlier discussion of self-averaging and spin symmetry, this determines the required $N_\eta$ in each setting to bound the stochastic density error. The resulting time-cost ratios are summarized in Table~\ref{tab:perform}. Taking the non-spin case as a reference, the stochastic TDH time cost for collinear calculations is 1$\sim$2 $\times$ larger, depending on the majority-to-minority spin ratio. The noncollinear nonmagnetic case requires roughly twice the cost, while the noncollinear magnetic case ranges between 2$\sim$4 $\times$ the non-spin baseline. We will demonstrate this observation in the next section. 

This increase is nevertheless modest compared to conventional deterministic $GW$, where a larger growth (4 $\times$) in computational effort is required due to the need for separate treatment of spin channels and expanded Hamiltonian matrix dimensions. In the stochastic framework, by contrast, the scaling remains linear, and the time cost increases by only about a factor of two in common cases such as spin-collinear and nonmagnetic SOC calculations.

\begin{table}
\caption{\label{tab:perform} Expected stochstic TDH time cost ratio}

\begin{tabular}{cccccccc}
  \hline
 \hline
& Scalar & Coll & NC\,+TR & NC\,–TR \\ \hline
$N_\eta$ & 1 & 1$\sim$2 & 1 & 1$\sim$2 \\ \hline
$\phi$ size & 1 & 1 & 2 & 2 \\ \hline
Time & 1 & 1$\sim$2 & 2 & 2$\sim$4 \\ 
 \hline
 \hline
\end{tabular}

\end{table}

\section{Simulation and discussion}

In this section, we present numerical simulations that verify the spin-resolved stochastic $GW$ formalism and analyze its scaling behavior in realistic settings. The simulations include both finite and periodic systems with strong SOC or ferromagnetic order. We first describe the computational setup and then compare the performance of the stochastic implementation in non-spin and spin-polarized cases.

The calculations were carried out on the Perlmutter supercomputer at NERSC. Each CPU node is equipped with two AMD Milan processors (64 cores total) and 512 GB of memory.  The implementation is trivially parallelized over stochastic orbitals and time steps, allowing efficient scaling across multiple nodes as in Ref.~\cite{vlvcek2018swift}. For molecules, the Martyna-Tuckerman approach is used to avoid periodic images~\cite{martyna1999reciprocal}. We apply Coulomb-interaction truncation along $z$ and 2D periodic boundary conditions for 2D systems~\cite{rozzi2006exact,brooks2020stochastic}.


\subsection{Collinear magnetic systems}

\begin{figure}
\includegraphics[scale=0.8]{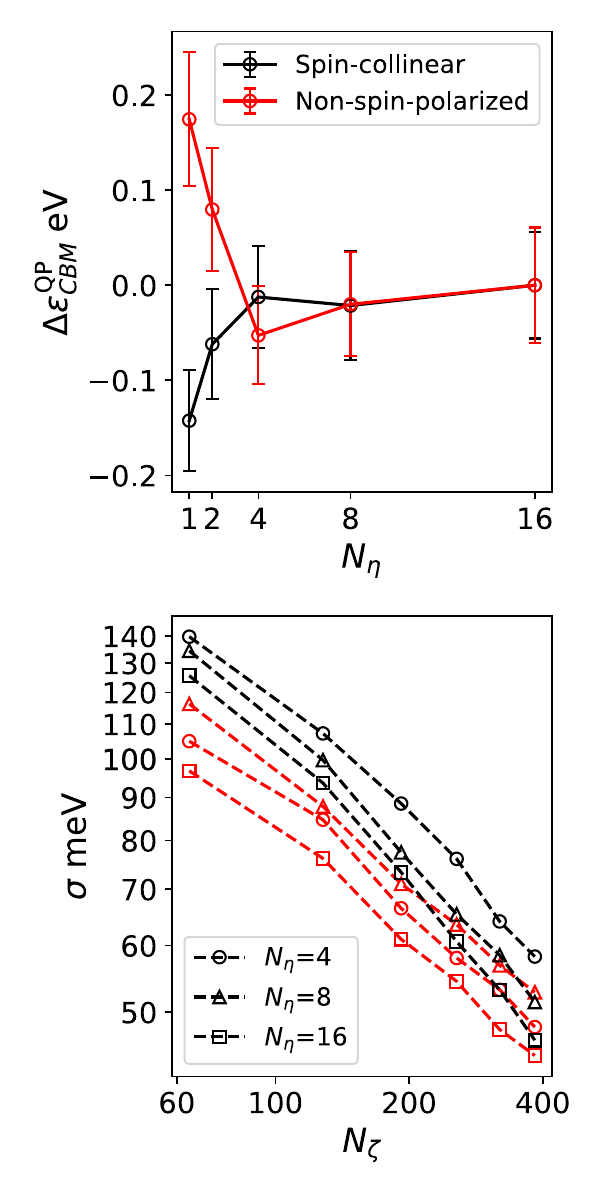}
\caption{\label{fig:cri3} Top: The CBM $E_{qp}$ of monolayer CrI$_3$ $4\times4$ supercell convergence test with increasing $N_\eta$. black: collinear spin polarized; red: non-spin-polarized; Bottom: the stochastic error convergence with increasing Monte Carlo samplings  $N_\zeta$ in log scale.} 
\end{figure}

For spin-collinear configurations, we verify the spin-resolved stochastic $GW$ method using the prototypical ferromagnetic monolayer CrI$_3$. It is a van der Waals layered material that hosts long-range ferromagnetic order down to the monolayer limit, making it a cornerstone of two-dimensional magnetism research~\cite{huang2017layer,gong2017discovery}. Its relatively simple crystal structure and Ising-like magnetic anisotropy provide a well-defined platform for testing electronic-structure methods. Our aim here is to demonstrate that the stochastic $GW$ framework accurately captures quasiparticle energies in such 2D ferromagnets, thereby establishing its applicability to spin-polarized layered systems. 


We use a dense real-space grid with grid intervals around 0.3 a.u. The Fritz-Haber-Institute pseudopotentials~\cite{fuchs1999ab} with local density approximation (LDA) are used. The convergence of the $4\times 4$ calculation depending on the $N_\eta$ is shown in Figure~\ref{fig:cri3} (top). The results show that the required $N_\eta$ for collinear and non-spin systems is therefore comparable. Only 4$\sim$8 stochastic orbitals $N_\eta$ (2$\sim$4 per spin channel) are sufficient to obtain accurate quasiparticle energies $E_{qp}$ within a 0.05 eV error bar at CBM. This finding is consistent with the performance analysis of the stochastic TDH approach that collinear s$GW$ requires almost the same number of $N_\eta$ sampling with no spin case. Figure~\ref{fig:cri3} (bottom) further confirms the expected $\sqrt{N_\zeta}$ scaling of stochastic errors. Then we perform s$GW$ calculation for CrI$_3$ from a $1\times1$ to a $6\times6$ supercell, involving up to thousands of electrons. The band gap results are shown in Table~\ref{tab:cri3}. 

These results demonstrate both the reliability of the chosen $N_{\eta}$ and the scalability of the stochastic $GW$ method for increasingly large system sizes, providing a practical balance between accuracy and efficiency.
 
\begin{table}
\caption{\label{tab:cri3} s$GW$ results for band gap of monolayer feromagnetic CrI$_3$. The calculations used $N_\zeta$= 160, $N_\eta$= 8.}

\begin{tabular}{cccccccc}
  \hline
 \hline
 Supercell  &$N_e$ &$N_g$&  gap (eV)  \\ \hline
 1x1   &   108& 64x32x128   &4.49$\pm$0.10 \\ \hline
 4x4   & 1728 & 256x128x128   &3.20$\pm$0.07 \\ \hline 
 6x6   & 2888 & 384x192x128  &3.10$\pm$0.09 \\ 
  \hline
 \hline
\end{tabular}

\end{table}

\subsection{Noncollinear systems}


\subsubsection{Finite systems}

We validated the noncollinear s$GW$ implementation on two finite systems with strong spin–orbit coupling (SOC): the I$_2$ molecule and the Pb$_4$S$_4$ nanocrystal that are previously used in deterministic $GW$ benchmarks~\cite{scherpelz2016implementation}. To check robustness against numerical choices, we employed dense real-space grids ($dx=0.27$ a.u. for I$_2$; $dx=0.3$ a.u. for Pb$_4$S$_4$) and two different Perdew-Burke-Ernzerhof (PBE) functional scalar-relativistic (SR) and fully relativistic (FR) norm-conserving pseudopotentials, PseudoDojo~\cite{hamann2013optimized} and SG15~\cite{hamann2013optimized,schlipf2015optimization,scherpelz2016implementation}.

For I$_2$, we computed SR and FR quasiparticle energies using deterministic TDH propagation to build the screening (Table~\ref{tab:molecule}). Relative to SR $G_0W_0$, the FR treatment—which includes SOC and noncollinear spinor dynamics—enhances screening and systematically lowers the ionization energy, in agreement with deterministic FR $GW$ results reported in Ref.~\cite{scherpelz2016implementation}.

We then examined the Pb$_4$S$_4$ nanocrystal, which preserves the rock-salt motif of bulk PbS. The calculations show noticeable differences in the absolute QP energies that can be attributed to the use of different pseudopotential versions, which may introduce variations of up to a few hundred meV~\cite{scherpelz2016implementation}.  We also see the difference between using PseudoDojo and SG15 pseudopotentials in Table~\ref{tab:molecule} and~\ref{tab:pbs}. Further differences come from distinct real-space grid treatments and employing the real-time (time-domain) $GW$ builds screening via TDH propagation, in contrast to (truncated) summation over states methods~\cite{govoni2015large}. Nevertheless, we can reliably confirm the role of the relativistic effects. The FR–SR correction to the HOMO–LUMO gap is
$\Delta E_g$ around 0.87 to 0.90 eV regardless of the pseudopotential used (Table~\ref{tab:pbs}), and it is consistent with deterministic $G_0W_0$ benchmarks (0.88 eV)~\cite{scherpelz2016implementation}.

Overall, these molecular and nanocrystal tests show that noncollinear s$GW$ reproduces the key FR quasiparticle shifts relative to SR while maintaining controlled stochastic uncertainty. This establishes a reliable baseline for applying the method to larger spin-polarized materials where fully deterministic FR $GW$ becomes computationally prohibitive.

\begin{table}
\caption{\label{tab:molecule} s$GW$ results for HOMO QP energies (eV) of I$_2$ using different pseudopotentials. The calculations used dx = 0.27 a.u.,$N_g$ = 64$^3$, $N_\zeta$= 640 and deterministic TDH propagation.}

\begin{tabular}{cccccccc}
  \hline
 \hline
 Pseudopotential & SR G$_0$W$_0$ & FR G$_0$W$_0$  \\ \hline
PseudoDojo & 9.80 $\pm$ 0.05 & 9.44 $\pm$ 0.05  \\ \hline
SG15  & 10.23 $\pm$ 0.05 & 10.11 $\pm$ 0.05 \\ \hline
SG15 (ref.~\cite{scherpelz2016implementation})  & 10.38 & 10.12 \\ \hline
Exp. (ref.~\cite{scherpelz2016implementation}) &  9.35\\
 \hline
 \hline
\end{tabular}

\end{table}

\begin{table}
\caption{\label{tab:pbs} s$GW$ results for HOMO-LUMO gap $E_g$ (eV) of Pb$_4$S$_4$ using different pseudopotentials. The $\Delta E_g$ is the difference between the SR and FR gap. The calculations used dx = 0.30 a.u., $N_g$ = 64$^3$ ,$N_\zeta$= 1000, $N_\eta$= 16.}

\begin{tabular}{ccccccccc}
  \hline
 \hline
 Pseudopotential & $E_g$ SR & $E_g$ FR & $\Delta E_g$ \\ \hline
 PseudoDojo & 6.12 $\pm$ 0.05 &  5.22 $\pm$ 0.05 & 0.9  \\ \hline 
 SG15 & 6.24 $\pm$ 0.05 & 5.37 $\pm$ 0.05  & 0.87  \\ \hline 
  SG15 (ref.~\cite{scherpelz2016implementation}) & 6.43 & 5.55 & 0.88  \\ 
   \hline
 \hline
\end{tabular}

\end{table} 

\subsubsection{Periodic system}

For the periodic system, we verify the spin-resolved s$GW$ method on bulk AlSb. A dense real-space grid with spacing $dx = 0.27$ a.u. is employed, using PseudoDojo~\cite{hamann2013optimized} PBE SR and FR pseudopotentials. The band gap calculations are shown in Table~\ref{tab:alsb} and are consistent with deterministic $GW$ results~\cite{scherpelz2016implementation}.

\begin{figure}
\includegraphics[scale=0.8]{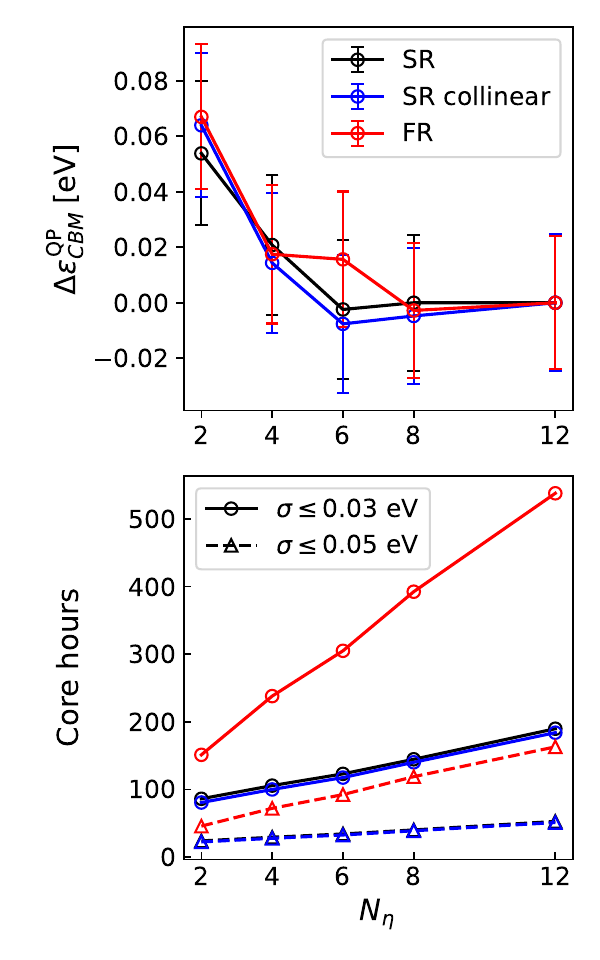}
\caption{\label{fig:alsb} Top: The CBM $E_{qp}$ of AlSb $2\times2\times2$ supercell convergence test with increasing $N_\eta$. Bottom: The core hours time cost of different $N_\eta$. Black: scalar relativistic; blue: collinear spin with SR pseudopotentials; red: fully relativistic} 
\end{figure}

We examine the convergence with respect to $N_\eta$, as shown in Figure~\ref{fig:alsb}. Both SR and FR pseudopotentials exhibit similar convergence, and $N_\eta = 8$ is sufficient to obtain an accurate CBM quasiparticle energy $E_{qp}$ as shown in Figure~\ref{fig:alsb} (top). This confirms the stochastic TDH performance analysis: noncollinear nonmagnetic systems require a comparable $N_\eta$ to achieve the same error level as collinear systems.
The corresponding total core-hour costs are presented in Figure~\ref{fig:alsb} (bottom). In the noncollinear case, the computational effort increases for two reasons: (i) the FFTs must be performed on spinor wavefunctions, effectively doubling their cost, and (ii) the SOC nonlocal pseudopotential terms introduce additional overhead. As a result, the runtime is roughly three times that of the spin-unpolarized case, rather than just twice. Nonetheless, this is still significantly more efficient than deterministic $GW$, where achieving a similar accuracy would require a factor of four increase in additional cost. Further, note that this is a cost increase relative to the spin-degenerate calculation. In the s$GW$ calculation, it scales merely linearly with a small prefactor, compared to the conventional calculations. Thus, the reported calculations are orders of magnitude less expensive than those obtained by a conventional route. This suggests that noncollinear s$GW$ preserves the favorable linear-scaling behavior observed in spin-unpolarized systems, consistent with the analysis in Refs.\cite{Neuhauser2014,vlvcek2018swift}.

\begin{table}
\caption{\label{tab:alsb} s$GW$ results for band gap (eV) of $2\times 2\times 2$ supercell AlSb. The calculations used $dx = 0.27$ a.u., $N_g$ = 128$^3$, $N_\zeta$= 480, $N_\eta$= 8.}

\begin{tabular}{cccccccc}
 \hline
 \hline
   &SR &FR  \\ \hline
  This work &   1.95$\pm$0.03 & 1.69$\pm$0.03   \\ \hline
  ref.~\cite{scherpelz2016implementation}    &1.89 & 1.65 \\ \hline
  \hline
\end{tabular}

\end{table}

\section{Conclusion and outlook}
We have extended the s$GW$ method to treat all spin configurations while preserving linear scaling. For spin-collinear magnetic systems, the computational cost remains nearly the same as in the non-spin-polarized case. By introducing a complex stochastic basis, we showed that noncollinear spinors can be handled without bias and used to estimate the self-energy expectation values. For time-reversal-symmetric noncollinear systems, the required number of stochastic functions $N_{\eta}$ in the TDH stage is unchanged relative to collinear systems, while the computational cost increases only modestly—by a factor of around 2–3—due to spinor FFTs and spin–orbit nonlocal pseudopotentials. In contrast, fully deterministic $GW$ calculations display a steeper overhead, with costs rising by roughly a factor of 4 compared to the non-spin-polarized case. Thus, the stochastic approach yields accurate quasiparticle energies at only a fraction of the deterministic cost, enabling large-scale studies of magnetic and spin–orbit–coupled materials that would otherwise be computationally prohibitive.

In the future, this unified s$GW$ framework
will lay the foundation for systematically incorporating
spin–spin interactions and vertex corrections. Such extensions
will be crucial for accurately capturing excitonic
effects, correlated magnetism, and unconventional superconductivity
in real materials. Besides, AI-based solvers and learning modules offer a natural path to further accelerate stochastic propagations, particularly by exploiting low-rank structures and transferable features across materials~\cite{zhu2025predicting,bassi2024learning,reeves2023dynamic,
mejia2023stochastic}. These developments highlight the synergy between stochastic methods and data-driven approaches, opening the door to large-scale quasiparticle studies of large scale complex spin–orbit–coupled and magnetic materials that would otherwise be computationally prohibitive.

\begin{acknowledgement}

This work was supported by the DOE BES grant DE-SC0024987 to V.V. This research used resources of the National Energy Research Scientific Computing Center (NERSC), a U.S. Department of Energy Office of Science User Facility operated under Contract No. DE-AC02-05CH11231, under NERSC Award No. BES-ERCAP0032056.

\end{acknowledgement}

\bibliography{main}

\begin{mcitethebibliography}{42}
\bibitem[sup()]{supp}
An equivalent way to circumvent this problem is to separate the real and imaginary parts of the $\phi$. We have, $\langle \phi_{\sigma} | \Sigma(t) | \phi_{\sigma'} \rangle =\langle \phi_{\sigma} | \Sigma(t) | Re[(\phi_{\sigma'}] \rangle + i \langle \phi_{\sigma} | \Sigma(t) |Im[\phi_{\sigma'}] \rangle$. Then, follow the procedure above, we calculate the two expressions separately. They both have positive source charge $\bar{\zeta}(r\sigma)Re[\phi_\sigma(r)]$ and $\bar{\zeta}(r\sigma)Im[\phi_\sigma(r)]$. However, this method doubles the time of calculating the effective interaction.\relax
\mciteBstWouldAddEndPunctfalse
\mciteSetBstMidEndSepPunct{\mcitedefaultmidpunct}
{}{\mcitedefaultseppunct}\relax
\EndOfBibitem
\end{mcitethebibliography}

\end{document}